\documentclass[twocolumn]{aastex631}

\begin{document}

\title{JWST/NIRCam Narrowband Survey of Pa$\beta$ Emitters in the Spiderweb Protocluster at z=2.16}

\author[0000-0003-4442-2750]{Rhythm Shimakawa}
\affiliation{Waseda Institute for Advanced Study (WIAS), Waseda University, 1-21-1, Nishi-Waseda, Shinjuku, Tokyo, 169-0051, Japan}
\affiliation{Center for Data Science, Waseda University, 1-6-1, Nishi-Waseda, Shinjuku, Tokyo, 169-0051, Japan}
\email{rhythm.shimakawa@aoni.waseda.jp}

\author[0000-0002-5963-6850]{J. M. P\'{e}rez-Mart\'{i}nez}
\affiliation{Instituto de Astrof\'{\i}sica de Canarias, E-38205 La Laguna, Tenerife, Spain}
\affiliation{Universidad de La Laguna, Dpto. Astrof\'{i}sica, E-38206 La Laguna, Tenerife, Spain}

\author[0000-0001-7147-3575]{Helmut Dannerbauer}
\affiliation{Instituto de Astrof\'{\i}sica de Canarias, E-38205 La Laguna, Tenerife, Spain}
\affiliation{Universidad de La Laguna, Dpto. Astrof\'{i}sica, E-38206 La Laguna, Tenerife, Spain}

\author[0000-0002-0479-3699]{Yusei Koyama}
\affiliation{Subaru Telescope, National Astronomical Observatory of Japan, National Institutes of Natural Sciences \\
650 North A'ohoku Place, Hilo, HI 96720, USA}

\author[0000-0002-2993-1576]{Tadayuki Kodama}
\affiliation{Astronomical Institute, Tohoku University, 6-3, Aramaki, Aoba, Sendai, Miyagi 980-8578, Japan}

\author[0000-0003-4528-5639]{Pablo G. P\'{e}rez-Gonz\'{a}lez}
\affiliation{Centro de Astrobiolog\'{\i}a (CAB), CSIC-INTA, Ctra. de Ajalvir km 4, Torrej\'{\o}n de Ardoz, E-28850, Madrid, Spain}

\author{Chiara D'Eugenio}
\affiliation{Instituto de Astrof\'{\i}sica de Canarias, E-38205 La Laguna, Tenerife, Spain}
\affiliation{Universidad de La Laguna, Dpto. Astrof\'{i}sica, E-38206 La Laguna, Tenerife, Spain}

\author[0000-0001-5757-5719]{Yuheng Zhang}
\affiliation{Purple Mountain Observatory, Chinese Academy of Sciences, 10 Yuanhua Road, Nanjing, 210023, China}
\affiliation{School of Astronomy and Space Science, University of Science and Technology of China, Hefei, Anhui 230026, China}
\affiliation{Instituto de Astrof\'{\i}sica de Canarias, E-38205 La Laguna, Tenerife, Spain}
\affiliation{Universidad de La Laguna, Dpto. Astrof\'{i}sica, E-38206 La Laguna, Tenerife, Spain}

\author[0000-0001-7713-0434]{Abdurrahman Naufal}
\affiliation{Department of Astronomical Science, The Graduate University for Advanced Studies, 2-21-1 Osawa, Mitaka, Tokyo 181-8588, Japan}
\affiliation{National Astronomical Observatory of Japan, 2-21-1 Osawa, Mitaka, Tokyo 181-8588, Japan}

\author[0000-0002-9509-2774]{Kazuki Daikuhara}
\affiliation{Astronomical Institute, Tohoku University, 6-3, Aramaki, Aoba, Sendai, Miyagi 980-8578, Japan}



\begin{abstract}

We report the initial result of our Pa$\beta$ narrowband imaging on a protocluster with the JWST Near Infrared Camera (NIRCam).
As NIRCam enables deep narrowband imaging of rest-frame NIR lines at $z>1$, we target one of the most studied protoclusters, the Spiderweb protocluster at $z=2.16$, in which previous studies have confirmed more than a hundred member galaxies. 
The NIRCam F405N narrowband filter covers in Pa$\beta$ line the protocluster redshift given by known protocluster members, allowing the search for new member candidates.
The weight-corrected color--magnitude diagram obtained 57 sources showing narrowband excesses, 41 of which satisfy further color selection criteria for limiting the sample to Pa$\beta$ emitter candidates at $z\sim2.16$, and 24 of them do not have H$\alpha$ emitter counterparts.
The Pa$\beta$ emitter candidates appear to follow the spatial distribution of known protocluster members; however, follow-up spectroscopic confirmation is required.
Only 17 out of 58 known H$\alpha$-emitting cluster members are selected as Pa$\beta$ emitters in the current data, albeit the rest fall out of the narrowband selection owing to their small Pa$\beta$ equivalent widths.
We derive Pa$\beta$ luminosity function in the Spiderweb protocluster, showing a normalization density of $\log{\phi^\ast}=-2.53\pm0.15$ at a characteristic Pa$\beta$ luminosity of $\log{L^\ast}=42.33\pm0.17$.
Furthermore, we examine the possibility of detecting faint line emitters visible only in the narrow-band image, but find no promising candidates.

\end{abstract}





\section{Introduction} \label{s1}

Exploring various galaxy populations and uncovering their physical properties through multi-wavelengths across large-scale structures and cosmic time lead to a greater understanding of the diverse growth histories of galaxies and how environmental effects shape the galaxy assembly.
As reviewed by, e.g.,~\citealt{Shapley2011,Conroy2013,Carilli2013,ForsterSchreiber2020,Tacconi2020}, a great deal of deep and high-resolution observations has been conducted for galaxies in the distant universe.
However, until recent years, we had lacked such high-quality data of the rest-frame near-infrared (NIR) emissions in high-redshift galaxies owing to the absence of suitable instruments, and thus they were limited to galaxies at $z\lesssim1.5$ (e.g., \citealt{Alonso-Herrero2006,Calzetti2007,Calabro2019,Gimenez-Arteaga2022}), except in the case of lensed galaxies \citep{Papovich2009}. 
Nevertheless, it is essential to investigate galaxies particularly at $z=$ 2--4, as during this redshift interval, known as the cosmic noon, the universe reached the peak epoch of cosmic star formation rate (SFR) density across cosmic time \citep{Madau1998,Lilly1999}.

With the advent of the James Webb Space Telescope (JWST), this situation has changed drastically; its powerful NIR instruments enable unprecedented depth in the rest-frame NIR observation for galaxies at $z=$ 1--3, with a spatial resolution of only $\sim0.1$ arcsec for the first time. 
A noteworthy aspect of the observational capabilities is the equipment with four narrowband filters (F323N/F405N/F466N/F470N) on the Near Infrared Camera (NIRCam; \citealt{Rieke2005}), enabling deep narrowband mapping of Pa$\alpha$ (or Pa$\beta$) lines at redshifts of $0.73/1.16/1.48/1.51$ (or $1.52/2.16/2.63/2.67$), respectively.
As Paschen lines are located in the NIR regime, they can be taken as nearly unbiased tracers of current star formation \citep{Kennicutt1998b,Calzetti2007}.
Indeed, new studies have begun investigating Pashcen lines of galaxies at $z>1$ with JWST. 
For instance, \citet{Neufeld2024} have revisited the star-forming main sequence using Pa$\alpha$ line and rest-frame NIR photometry, which shows that its scatter of $\sim0.29$ dex is consistent with those of previous studies based on the rest-frame optical lines or SED fitting.
\citet{Reddy2023} have suggested the possible presence of star formation not captured by Balmer lines, exemplifying the scientific importance of Paschen line observations.

One of the NIRCam narrowband filters can cover Pa$\beta$ lines of galaxies in one of the most studied galaxy protoclusters, the Spiderweb protocluster \citep{Carilli1997,Pentericci1997,Pentericci2000,Kurk2000}, associated with the Spiderweb radio galaxy, PKS~1138$-$262 at $z=2.16$ \citep{Bolton1979,Roettgering1994,Roettgering1997,vanOjik1995}. 
Beginning with the first discovery of significant overdensity of Ly$\alpha$ emitters \citep{Kurk2000,Pentericci2000}, more than a hundred protocluster member galaxies have now been identified across various wavelength regimes (e.g., \citealt{Kurk2004a,Kurk2004b,Croft2005,Doherty2010,Tanaka2010,Tanaka2013,Koyama2013a,Shimakawa2014,Shimakawa2018b,Jin2021,Perez-Martinez2023}, and see, e.g., \citealt{Shimakawa2018b,Tozzi2022a} for overviews of previous surveys).
The protocluster core has recently been confirmed by the thermal Sunyaev–Zeldovich effect \citep{DiMascolo2023}. 
They report that the halo mass is estimated to be $\mathrm{M_{500}}=3.46\times10^{13}$ M$_\odot$, representing progenitors of galaxy clusters with $\mathrm{M_{500}}=(5.6\mathrm{-}8.8)\times10^{14}$ M$_\odot$ at $z=0$ (see also \citealt{Shimakawa2014,Tozzi2022b,Lepore2023}).
\citet{Lepore2023} measured a temperature profile inside the protocluster core for the first time and then suggested the presence of a strong cool flow with 250--1000 M$_\odot$yr$^{-1}$ and an energy budget of ongoing feedback power beyond $\sim10^{43}$ erg~s$^{-1}$, which is primarily driven by the radio galaxy.
In fact, the abrupt changes owing to the synergistic effect of gas feeding and feedback mechanisms are expected in protocluster halos in this period, according to cosmological simulations, such as cold streams in hot halos \citep{Dekel2009,Suresh2019,Stern2020} and preheating driven by AGN feedback \citep{Pratt2010,Chaudhuri2013,Iqbal2017,Kooistra2022}.
These observations and simulations all suggest that the Spiderweb protocluster is rapidly transitioning to the bonafide cluster. 
Hence, it provides important insights into galaxy cluster formation.

Motivated by such backgrounds, we conduct Pa$\beta$ narrowband imaging towards the Spiderweb protocluster with JWST/NIRCam (Cycle~1 GO~1572, P.I. Dannerbauer and Koyama).
The deep narrowband (F405N) data provide us with a new window to search for star-forming galaxies and active galactic nuclei (AGNs) in Pa$\beta$ line at $\lambda=1.28$ $\mu$m in the Spiderweb protocluster.
Therefore, the main scope of this study, as part of a series of early reports with P\'{e}rez-Mart\'{\i}nez et al. (2024), is to explore this classic protocluster from yet another perspective of Pa$\beta$ line and to offer unique member candidates. 
This paper is organized as follows: we first summarize the NIRCam dataset (Section~\ref{s2}), then we demonstrate the narrowband technique and color--color selection for selecting candidates of protocluster members (Section~\ref{s3}). 
Here, we utilize the archival H$\alpha$ emitters (HAEs) from \citet{Shimakawa2018b,Shimakawa2024} for the purpose of validation. 
Subsequently, we discuss the spatial distribution of the obtained candidates of Pa$\beta$ emitters associated with the Spiderweb protocluster, and measure the Pa$\beta$ luminosity function in the survey field (Section~\ref{s4}).
We also discuss the possible detection of emission-line-dominated candidates detected only in the narrowband image.
Finally, we summarize the entire flow of the paper and the obtained results in Section~\ref{s5}.

We assume a flat lambda cold dark matter model with $h=0.693$ and $\Omega_M=0.286$, consistent with those obtained from the WMAP nine-year data \citep{Hinshaw2013}.
We use the \citet{Chabrier2003} stellar initial mass function (IMF) and the AB magnitude system \citep{Oke1983}.
We assume a positive equivalent width (EW) for emission lines unless otherwise noted.


\section{Data and sample} \label{s2}

\subsection{NIRCam imaging} \label{s21}

\subsubsection{Observation} \label{s211}

\begin{figure}
\centering
\includegraphics[width=8.5cm]{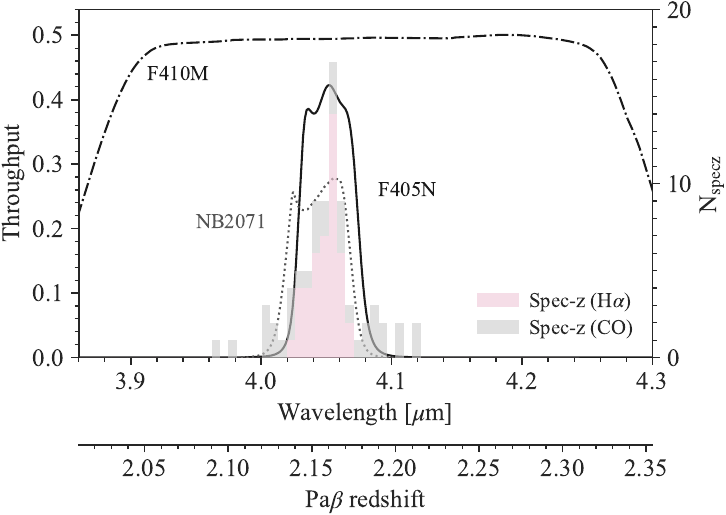}
\caption{
Filter response curves and redshift distribution of spectroscopically confirmed members in the Spiderweb protocluster.
The dotted, solid, and dash-dot lines indicate the filter throughput of the MOIRCS/NB2071, NIRCam/F405N, and F410M bands, respectively.
Spectroscopic confirmations with H$\alpha$ and CO(1--0) lines are shown separately by pink and gray (e.g., \citealt{Shimakawa2014,Jin2021,Perez-Martinez2023}).
}
\label{fig1}
\end{figure}

We conduct narrowband imaging on the Spiderweb protocluster (Cycle1 GO~1572, P.I. Dannerbauer and Koyama) with the F405N and F410M filters (in parallel, F115W and F182M filters on the blue channel) on the JWST Near Infrared Camera (NIRCam, \citealt{Rieke2005}).
The NIRCam F405N filter ($\lambda_\mathrm{center}=4.055$ $\mu$m, FWHM = 0.046 $\mu$m) covers a Pa$\beta$ redshift range of $z=2.165\pm0.018$, which covers well the protocluster redshift and largely overlaps the redshift coverage of $z=2.156\pm0.025$ by the previous H$\alpha$ imaging (Fig.~\ref{fig1}).
The spatial resolution of 0.14 arcsec in full-width-at-half-maximum (FWHM) in the F405N band is comparable to those of HST/ACS (FWHM = 0.11 arcsec; e.g., \citealt{Stevens2003}) and ALMA Band-7 data (FWHM = 0.16 arcsec; Koyama et al. in preparation). 
Additionally, the NIRCam F405N filter pairs with our companion filter, NB2071, installed on the Multi-Object Infrared Camera and Spectrograph (MOIRCS) on the Subaru Telescope \citep{Ichikawa2006,Suzuki2008}, allowing us to analyze Pa$\beta$ lines of the existing H$\alpha$ emitters in the Spiderweb protocluster \citep{Koyama2013a,Shimakawa2018b}, as reported in the companion paper (P\'{e}rez-Mart\'{\i}nez et al. 2024).

\begin{figure}
\centering
\includegraphics[width=8cm]{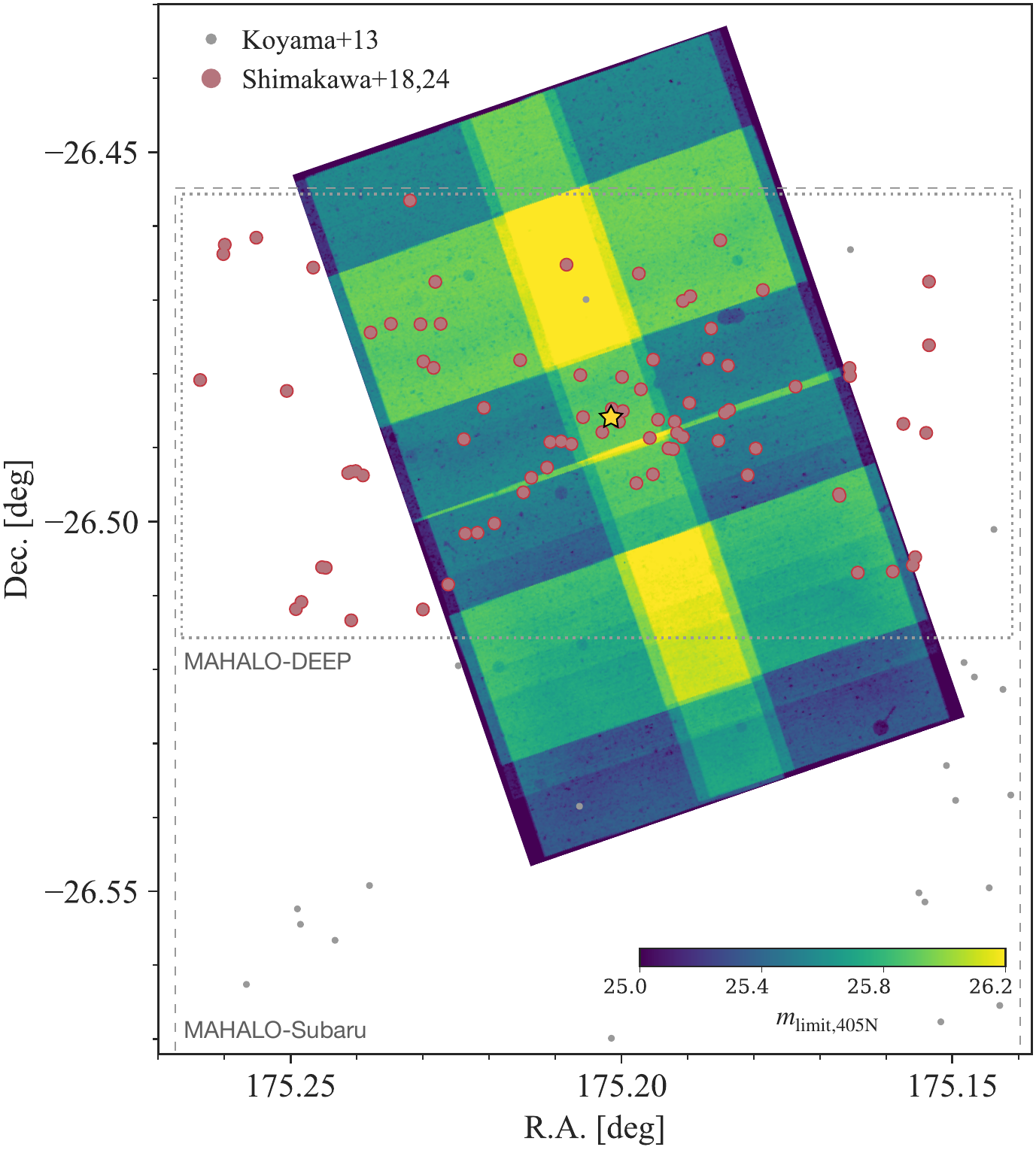}
\caption{
Spatial distribution of known HAEs from previous narrowband imaging. 
The red circles show the HAE samples in the deep field depicted by the dotted square (MAHALO-DEEP; \citealt{Shimakawa2018b,Shimakawa2024}), and the gray dots indicate additional HAEs from the original narrowband imaging covering two times wider as depicted by the dashed square (MAHALO-Subaru; \citealt{Koyama2013a}). 
The yellow star indicates the Spiderweb radio galaxy.
Color map corresponds to the imaging depth of the F405N image. 
}
\label{fig2}
\end{figure}

In the observing run, we set a primary dither pattern of {\tt FULLBOX 4TIGHT} of {\tt ROWS $\times$ 2} with 35~\% overlap and {\tt COLUMNS = 1} with 10 percent overlap, which can cover the $3^\prime\times6^\prime$ square field around the Spiderweb galaxy, which amounts to the total number of exposures $N_\mathrm{exp}=2-8$ per field (Fig.~\ref{fig2}). 
The {\tt FULLBOX TIGHT} dither is adopted to maximize observing efficiency. 
The maximum integrations reach 63 min in the F405N and F115W bands and 21 min in the F410M and F182M bands.
Also, the position angle is not fixed to allow greater flexibility in the observing schedule. 
Approximately 60\% of the survey area is covered by the H$\alpha$ narrowband imaging with Subaru/MOIRCS \citep{Koyama2013a,Shimakawa2018b} and the wide CO~(1--0) emitter survey with ATCA \citep{Jin2021}. 
In this work, we employ only H$\alpha$ (+ [N{\sc ii}])-emitting galaxies (HAEs) from \citet{Shimakawa2018b}, whose sky coordinates are available in \citet{Shimakawa2024}.
Of these, 58 HAEs are covered by the NIRCam survey area (Fig.~\ref{fig2}), and thus, they can be used to select Pa$\beta$ emitter candidates in this work, as shown in Section~\ref{s3}.

\subsubsection{Data reduction} \label{s212}

The raw frames are downloaded from the Barbara A. Mikulski Archive for Space Telescopes (MAST)\footnote{\dataset[10.17909/vx25-q902]{\doi{10.17909/vx25-q902}}}, processed by Science Data Processing (SDP) version 2023-1a, and post-processed with the JWST Science Calibration pipeline version 1.10.2 under the Calibration Reference Data System (CRDS) context 1140.pmap \citep{Bushouse2023}. 
Apart from the three official pipeline stages, we carry out several offline procedures to improve the calibration using the Rainbow JWST pipeline, described in \citet{Perez-Gonzalez2024}. 
In particular, $1/f$ noise\footnote{\url{https://jwst-docs.stsci.edu/known-issues-with-jwst-data/nircam-known-issues}} is removed, and the background is homogenized with a combination of row/column and box-median filtering. 
In both cases, sources are masked before the procedure using a mask produced with a dry run of the official pipeline, including a mask dilation, and by-hand masking of the brightest, most extended sources in the cluster (and some bright stars). 

The final mosaics are built with a 0.03 arcsec pixel scale, and their imaging qualities are evaluated in Section~\ref{s22}.
World Coordinate System is calibrated using the {\sc TweakReg} patch developed by CEERS collaboration \citep{Bagley2023} using as reference for the JWST data the F410M image (the deepest one), and then aligning this frame to the HST/ACS F814W image \citep{Miley2006}. 
For this task, we create a catalog containing sources within the overlapping area between the JWST and HST filters. 
We request our sources to fulfill $\mathrm{S/N>10}$ per pixel within a minimum area equivalent to twice the size of the point-spread-function (PSF) in the respective filters and remove objects with significant elongations or distortions which could introduce additional uncertainties in the determination of their positions.

\subsection{Source detection} \label{s22}

\begin{figure*}
\centering
\includegraphics[width=0.9\textwidth]{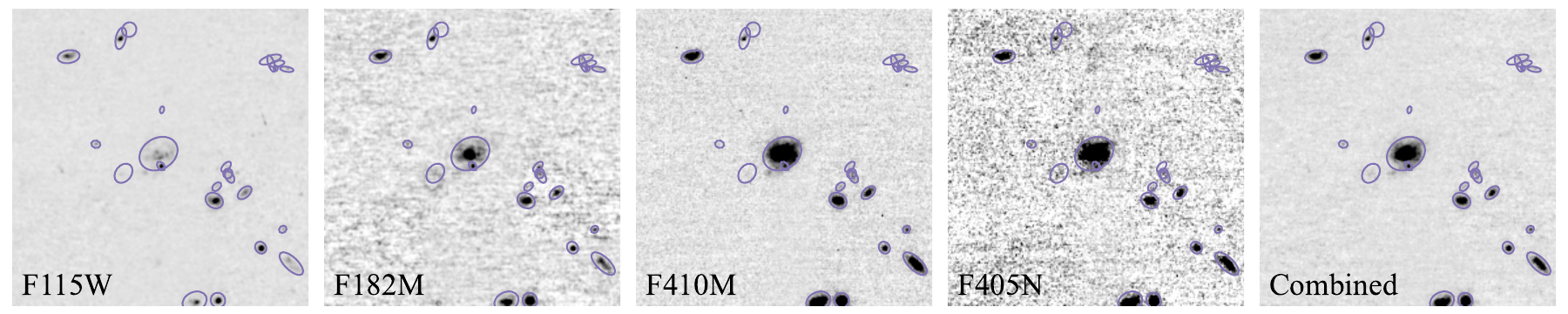}
\caption{
Example cutout images with $0.3\times0.3$ arcmin$^2$.
From the left to right, panels show the F115W, F182M, F410M, F405N band images, and the 4-band combined image, respectively, on the same flux scale in counts.
Here, the F115W and F182M images are smoothed by the Gaussian kernels to match their resolutions to the F410M and F405N images.
The magenta circles indicate Kron radius obtained from the combined image for each detection.
}
\label{fig3}
\end{figure*}

We conduct source detection and noise estimation based on the final mosaic images.
These tasks have to be done carefully because the imaging depth is not homogeneous across the survey field (Fig.~\ref{fig2}). 
In addition to the different number of exposures ($N_\mathrm{exp}=2-8$), many dips of the imaging depth exist in the fields even with the same net integration, owing to the sensitivity variance between detectors and within individual detectors.
These factors significantly affect the noise estimations, and hence the narrowband selection as described in the following (Section~\ref{s3}).
We are also concerned about a flux loss in the galaxy outskirts because diffuse components may be missed on the NIRCam images, especially in the F450N band, which is relatively shallower compared to the other medium or broad bands.
This may cause serious underestimations of Kron radii when measuring total flux densities.
Besides, the NIRCam data reduction pipeline still needs to be improved as it leaves some artifacts, such as snowballs, due to the impacts of cosmic rays\footnote{\url{https://jwst-docs.stsci.edu/data-artifacts-and-features/snowballs-and-shower-artifacts}}, critical for narrowband imaging requiring long exposure per frame.
Such artifacts may be present as false narrowband emitters (NBEs) if we do not clear them adequately (see also Section~\ref{s43}). 

Considering various concerns raised above, we adopt a median combined image of all four available bands (F115W + F182M + F410M + F405N) for the source extraction (Fig.~\ref{fig3}), enabling us to minimize the aforementioned issues that may happen when only the narrowband image is used.
For instance, the cosmic ray effect can be mitigated in the median stacked image, and its deeper imaging depth helps us determine the Kron radius relatively well, as we can reduce the risk of overlooking diffuse components. 
Instead, this decision causes another potential issue of ignoring emission-line dominated objects (so-called NB-only emitters), visible only in the narrowband image but not in the other bands (see, e.g., \citealt{Hayashi2016,Shimakawa2018a}).
However, determining whether such faint-end sources are real or not (e.g., noise and artifact) is challenging, given the current limitations of the JWST/NIRCam data reduction tool.
Therefore, these peculiar objects are discussed briefly in Section~\ref{s43}.

\begin{figure}
\centering
\includegraphics[width=8cm]{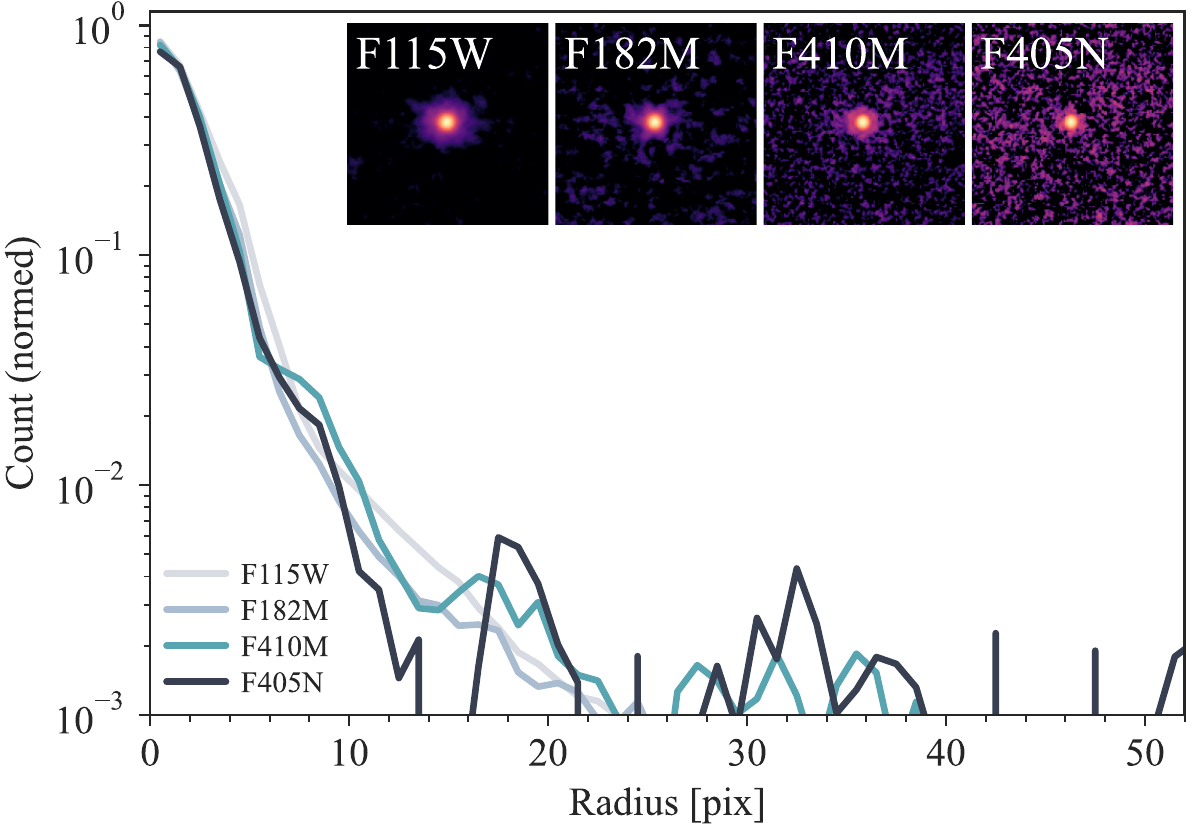}
\caption{
Radial profiles of composite point sources in the 4 photometric bands.
Pixel scale is 0.03 arcsec.
In order from the lightest color, the curves show those in the F115W, F182M, F410M, and F405N bands, respectively.
Their composite images are inset in the upper right.
We selected only 17 faint point sources at the forced positions based on the combined image because most of point sources are saturated in the F115W band.
}
\label{fig4}
\end{figure}

Cutout examples of reduced images can be found in Fig.~\ref{fig3}.
We combine all four available band images in the median after aligning their pixel coordinates and point source FWHMs.
Fig.~\ref{fig4} indicates radial profiles of composite images of 17-point sources in the four bands, showing that their radial profiles of agree well with each other.
Notably, we have selected only faint point sources that are not saturated in the F115W band. 
Note that PSF wings do not affect the narrowband selection as they are beyond the aperture radius of 0.15 arcsec or 5 pixels (Section~\ref{s31}).
In Fig.~\ref{fig4}, the point sources have been extracted at the forced pixel coordinates for all the imaging data. 
Therefore, the consistent radial profiles also indicates that their pixel coordinates are overall consistent with each other on the sub-pixel scale.
Then, we perform source detection using {\sc SExtractor} (version 2.19.5, \citealt{Bertin1996}) with detection parameters of {\tt DETECT\_MINAREA} $=25$, {\tt DETECT\_THRESH} $=1.5$, {\tt ANALYSIS\_THRESH} $=1.5$, where we have determined these parameters through trial and error to avoid detecting visible artifacts. 
We use default deblending parameters in {\tt DEBLEND\_NTHRESH} $=32$ and {\tt DEBLEND\_MINCONT} $=0.005$.
After manually masking bright stars and their diffraction spikes, we run the source extractor in the dual imaging mode, weighting through a weight map of the combined image.
Consequently, 16,444 sources are detected, including noisy objects.
Examples of source detection and their obtained Kron radii in each image can be found in Fig.~\ref{fig3}. 

\begin{figure}
\centering
\includegraphics[width=8cm]{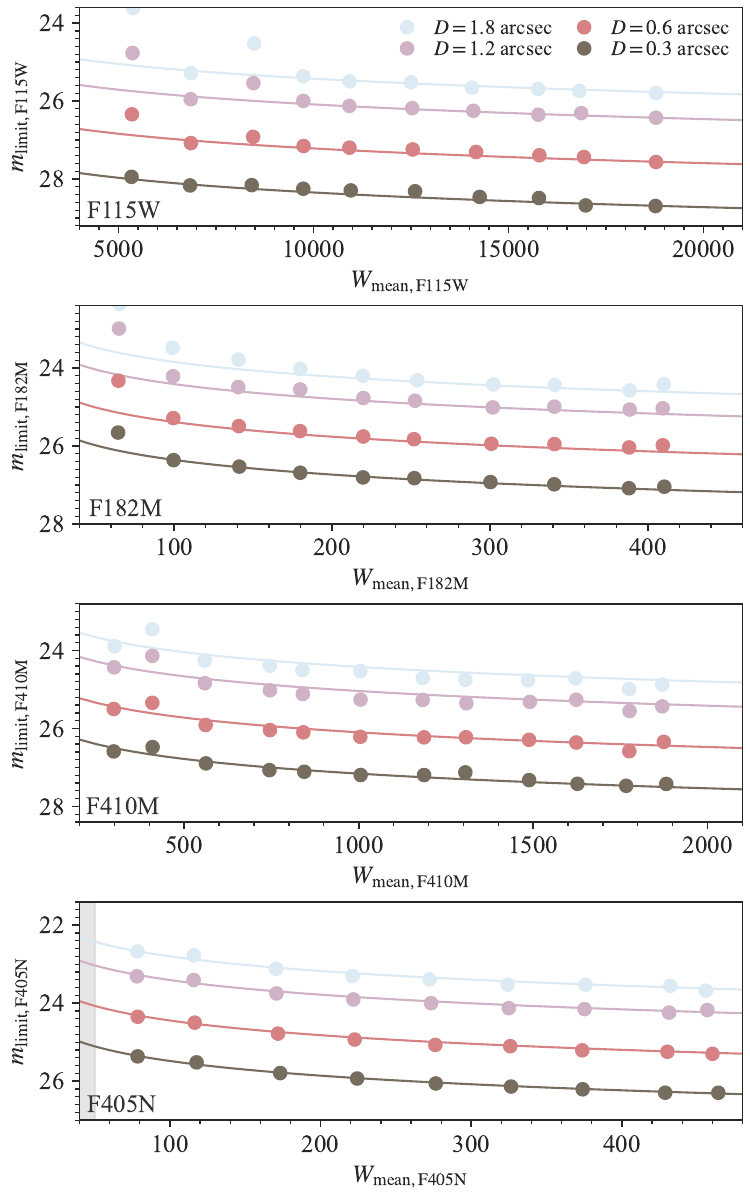}
\caption{
Limiting magnitude ($5\sigma$) as a function of the mean weight value in different aperture diameters.
From the top, each panel shows the limiting magnitude in the F115W, F182M, F410M, and F405N band, respectively.
The black, red, magenta, and cyan circles show some cases of derived limiting magnitudes in aperture diameters of 0.3, 0.9, 1.2, and 1.8 arcsec, respectively, and their best-fit error functions are depicted by corresponding color curves (Eq.~\ref{eq1}-\ref{eq4}).
We mask noisy edges in the F405N image in the source extraction, shown by the gray filled area in the bottom panel.
}
\label{fig5}
\end{figure}

We employ a commonly-used technique for noise estimation based on random aperture photometry but with various aperture diameters while measuring the mean weight values within the aperture areas. 
We then obtain the best-fit error functions, $\sigma(D,W)$, given circularized diameter and weight values using {\sc photutils} package (version~1.10.0, \citealt{Bradley2023}), enabling the estimation of photometric errors for all detected sources. 
Here, $D$ is a circularized aperture diameter ($D=0.3$ arcsec is applied in the narrowband selection), and $W$ is the mean weight value within a corresponding aperture area in the weight map. 
We distribute empty apertures ($N\sim30,000$ in each radius) with 0.15--1.20 arcsec radii variable by 0.15 arcsec on blank fields in each image. 
We then derived limiting magnitudes and the best-fit functions in the four NIRCam imaging data (Fig.~\ref{fig5}).
Consequently, we confirm that the sky errors decrease with the square root of weight values, $W^{-0.5}_\mathrm{band}$, which is natural by definition. 
They also increase with increasing aperture diameter by $D^{\sim1.4}$ owing to partial pixel-to-pixel correlations.
The obtained error function in each band can be described as follows:
\begin{eqnarray}
\sigma(D,W)_\mathrm{F115W} &=& 15.760~W^{-0.5}_\mathrm{F115W}\times (D/5)^{1.501}, \label{eq1} \\
\sigma(D,W)_\mathrm{F182M} &=& ~~9.825~W^{-0.5}_\mathrm{F182M}\times (D/5)^{1.292}, \label{eq2} \\
\sigma(D,W)_\mathrm{F410M} &=& 14.874~W^{-0.5}_\mathrm{F410M}\times (D/5)^{1.408}, \label{eq3} \\
\sigma(D,W)_\mathrm{F405N} &=& 22.509~W^{-0.5}_\mathrm{F405N}\times (D/5)^{1.380}. \label{eq4}
\end{eqnarray}
Uncertainties of flux densities in fixed aperture and Kron photometry for each source are estimated based on these prescriptions throughout the study by adopting their circularized diameters.
Five sigma limiting magnitudes in Fig.~\ref{fig5} are derived through $m_\mathrm{limit}\equiv\mathrm{ZP}-2.5\log(5\sigma(D,W))$, where $\mathrm{ZP}$ is the zero-point magnitude.
We adopt a zero-point magnitude of 28.065 mag for all bands except for the narrowband (F405N) image, including a small zero-point correction of $+0.03$ mag to fit a zero-point offset ($=0.00$ mag) between F410M and F405N (see also Section~\ref{s32}).
For the obtained images, their limiting magnitudes are consistent with those expected from the JWST Exposure Time Calculator \citep{Pontoppidan2016}. 
However, the different net integrations (up to $\sim0.75$ mag given $N_\mathrm{exp}=$2--8) and sensitivity variances within individual detectors lead to the differential of imaging depths up to $\sim$ one mag, depending on the source position.

\section{Narrowband selection} \label{s3}

\subsection{Color--magnitude selection} \label{s31}

To search for Pa$\beta$ emitter (PBE) candidates at $z\sim2.16$, we perform narrowband selection similarly to previous approaches (e.g., \citealt{Bunker1995,Sobral2009,Sobral2013,Hayashi2010,Matthee2017,Shimakawa2018a}) as the following prescriptions:
\begin{eqnarray}
\mathrm{BB}-\mathrm{NB} &>& -2.5\log(1-\Sigma\delta10^{-0.4(\mathrm{ZP}-\mathrm{NB})}),\label{eq5} \\
\mathrm{BB}-\mathrm{NB} &>& 0.124,\label{eq6}
\end{eqnarray}
where $\Sigma$ is a sigma confidence level ($\Sigma=3$ is adopted in this work) of narrowband (F405N) excess relative to F410M in a fixed aperture photometry of 0.3 arcsec diameter, and $\delta$ represents the combined $1\sigma$ background noise in the NB (F405N) and BB (F410M) bands, defined by $\delta^2=\sigma(D,W)^2_\mathrm{NB}+\sigma(D,W)^2_\mathrm{BB}$ (see Eq.~\ref{eq3}-~\ref{eq4}). 
The latter color threshold (Eq.~\ref{eq6}) is a selection cut for the EW limit in the narrowband flux ($=20$ \AA\ in the rest frame at $z=2.16$), which is comparable to previous narrowband surveys.

\begin{figure}
\centering
\includegraphics[width=8.5cm]{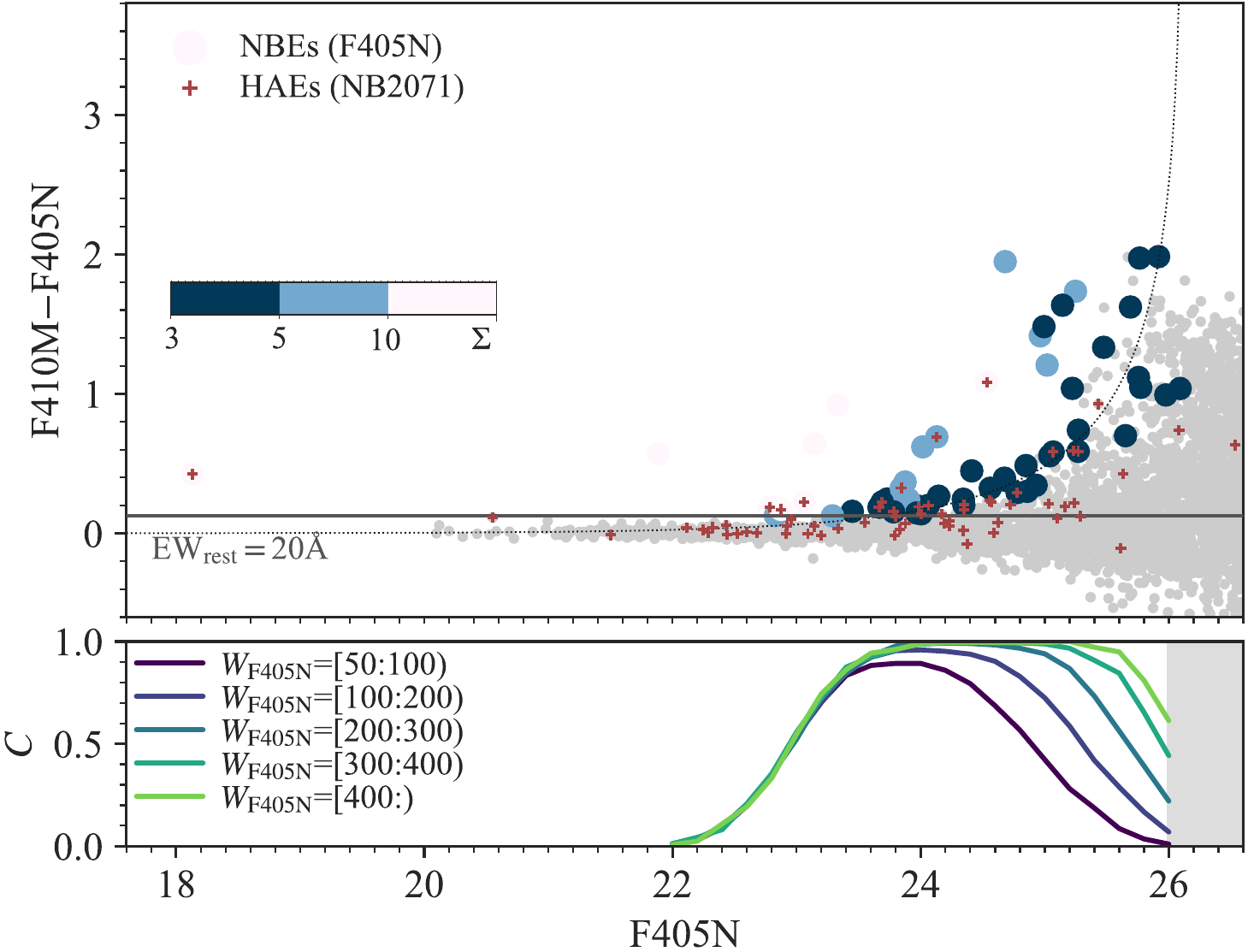}
\caption{
Top: Color--magnitude diagram for the NBE selection (see Eq.~\ref{eq5}-\ref{eq6}).
The gray dots show all sources in the F405N band, and the colored circles represent the emitter candidates above $3\sigma$ confidence levels, where symbol colors indicate significance levels of the narrowband excess (see the color bar) and the mean color threshold is depicted by the dotted curve.
The red dots mark known HAEs \citep[table~A]{Shimakawa2024}.
The black horizontal line depicts the EW cut ($=20$ \AA\ in the rest frame).
Bottom: Selection completeness, $C$ (recovery rate of mock NBEs) as a function of F405N (see Section~\ref{s31} for more details).
Line colors indicate the completeness in different weight values (from the darker color, $W_\mathrm{F405N}=[50:100),[100:200),[200:300),[300:400),[400:)$, where higher weights mean deeper as in Eq.~\ref{eq4} and Fig.~\ref{fig5}).
}
\label{fig6}
\end{figure}

We perform narrowband selection based on the obtained noise functions, as described in Section~\ref{s22}.
The resultant color--magnitude diagram is represented in Fig.~\ref{fig6}, where we plot 6,384 sources detected in the F410M band at greater than $2\sigma$ levels. 
The typical flux limit ($\Sigma=3$) corresponds to $\sim2\times10^{-18}$ erg~s$^{-1}$cm$^{-2}$. 
As expected, this is comparable with the H$\alpha$+[N{\sc ii}] flux limit of $\sim3\times10^{-17}$ erg~s$^{-1}$cm$^{-2}$ in our previous narrowband imaging with Subaru/MOIRCS \citep{Shimakawa2018b}, given the dust-free line ratio of H$\alpha$/Pa$\beta=17.5$ in the case B recombination \citep{Luridiana2015}.
After careful visual confirmation, we obtain 57 emitter candidates showing color excesses above $\Sigma=3$ 
(Eq.~\ref{eq5}) and the rest-frame EW $>20$ \AA\ (Eq.~\ref{eq6}), 17 of which are successfully matched to the known HAEs within 0.5 arcsec \citep{Shimakawa2018b}.

Notably, although most known HAEs are expected to have sufficiently bright Pa$\beta$ line fluxes, only 17 out of 58 HAEs show significant narrowband excesses in the F405N band (Fig.~\ref{fig6}). 
We confirm that this trend is particularly evident in the massive end with M$_\star>4\times10^{10}$ M$_\odot$ \citep[table~A1]{Shimakawa2024}, where only the Spiderweb radio galaxy meets the selection criteria (1/13).
The major reason for missing HAEs in Pa$\beta$ lines is a selection bias due to the small aperture diameter of 0.3 arcsec and the EW cut adopted for the selection. 
This aperture size has been set to maximize the NIRCam's great sensitivity to compact objects and minimize the effects of nearby cosmic ray pixels; however, it can track only central regions within a 1.27 kpc radius and cannot trace outer star formation.
Thus, massive star-forming galaxies forming central bulges cannot be trapped by the narrowband selection due to the relatively small EWs in the galaxy centers.
Indeed, the forthcoming paper (Shimakawa et al. in preparation) confirm the typical rest-frame EW$_\mathrm{Pa\beta}$ $\sim10$ \AA\ of such massive HAEs in the aperture area, which is lower than the selection threshold of 20 \AA, according to the composite image of massive HAEs with M$_\star>2\times10^{10}$ M$_\odot$.
Line flux ratios for individual HAEs are examined in detail by P\'{e}rez-Mart\'{\i}nez et al. (2024).

Furthermore, we examine the recovery rates of the narrowband selection through Monte Carlo simulation similarly to previous studies \citep{Sobral2009,Sobral2013,Shimakawa2018a}. 
The calculation is performed as follows. 
First, 200 PSF images made from composite images of point sources are embedded at random positions in the F405N and F410M images. 
Here, for a given F405N magnitude, F410M magnitudes are determined following a regression line between F405N and F410M for the selected NBEs, with a scatter of 0.1 mag in view of color (EW) variations.
We run 50 iterations in the Monte Carlo simulation for a given F405N magnitude in a step of 0.2 mag at 22--26 mag; hence, 10,000 point sources are embedded in each magnitude bin. 
We then evaluate the recovery rate (known as selection completeness, $C$) by applying narrowband selection (Eq.~\ref{eq5}-\ref{eq6}) for the embedded sources.
The obtained selection completeness are summarized in Fig.~\ref{fig6}, which plots completeness values in different weight intervals considering the imaging depth variation.
It shows that there is a significant variance of the selection completeness depending on the source location: the completeness still reaches approximately 60\% at the faint end if the source is located in the deepest region ($W_\mathrm{F405N}\geq400$), while dropping to almost zero in the shallow regions ($W_\mathrm{F405N}<200$).
We can also see that the selection completeness decreases towards the bright end (F405N $<23$ mag),  regardless of weight values, owing to lower EWs of NBEs.
We adopt the obtained completeness values when measuring Pa$\beta$ luminosity function in Section~\ref{s42}.

\subsection{Color--color selection} \label{s32}

This section examines which emission lines may affect our narrowband selection, and then tries to select more plausible candidates associated with the Spiderweb protocluster using color--color selection.
Even though the selected NBEs are more likely to be Pa$\beta\lambda12822$ emitters at $z=2.2$ compared with those in the random field because of their significant over-densities (e.g., \citealt[$\delta\sim10$]{Shimakawa2018b}); various strong emission lines at different redshifts may contribute to the emitter sample. 
Specifically, we may have line emitter contaminants, such as PAH~$\lambda3.3\mu$m at $z=0.2$ or Pa$\alpha$ at $z=1.2$ in the foreground, and H$\alpha$ at $z=5.2$ or [O{\sc iii}] doublet at $z\sim7.1$ in the background. 

\begin{figure}
\centering
\includegraphics[width=8cm]{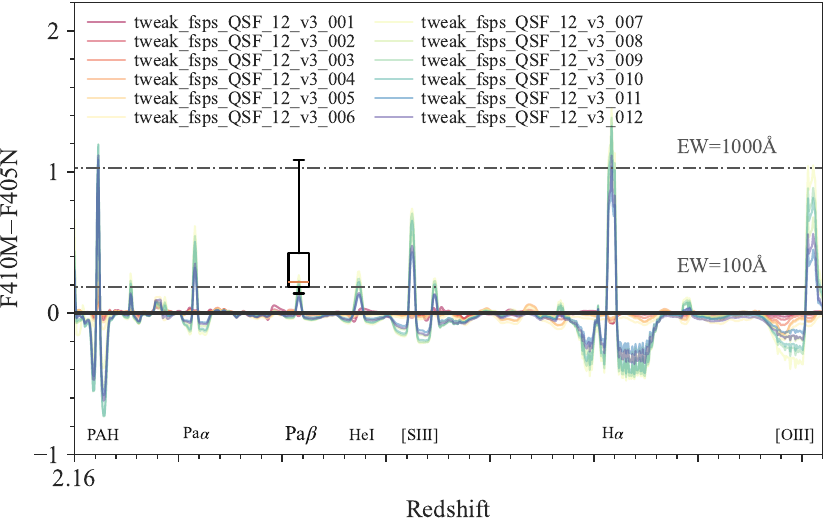}
\caption{
Emission line contributions to the narrowband magnitude (F410M$-$F405N) depending on galaxy redshifts.
The box plot shows the 25--75th percentiles of those of 17 PBEs confirmed with HAE counterparts and its error-bar depicts the whole range.
Colored lines indicate the color term values of twelve model templates of star-forming and quiescent galaxies ({\tt tweak\_fsps\_QSF\_12\_v3}, see legends and text) from {\sc eazy} \citep{Brammer2008}.
Strong emission lines are appended in the figure; however, it should be noted that the emission line models are limited and do not comprehensively reproduce narrowband excesses (see text).
}
\label{fig7}
\end{figure}

Fig.~\ref{fig7} demonstrates which emission lines significantly impact narrowband excess (F410M$-$F405N) depending on the source redshift, based on SED templates of star-forming and quiescent galaxies from a program termed Easy and Accurate Redshifts from Yale ({\sc eazy}, \citealt{Brammer2008}).
We adopt a set of twelve standard templates ({\tt tweak\_fsps\_QSF\_12\_v3}) from available models in the software program, showing that PAH~$\lambda3.3\mu$m at $z=0.2$, Pa$\alpha\lambda18756$ at $z=1.2$, [S{\sc iii}]$\lambda\lambda$9071,9533 doublet at $z\sim3.3$, H$\alpha\lambda6565$ at $z=5.2$, or [O{\sc iii}]$\lambda$4960,5008 doublet at $z\sim7.1$, would greatly contribute to the narrowband selection. 
It should be noted here that narrowband excesses vary more widely with galaxy properties than indicated in the figure (e.g., \citealt{Sobral2013}).
Indeed, most of confirmed PBEs with HAE counterparts show higher narrowband excesses (EW~$\gtrsim100$~\AA) than the model templates.
On the other hand, for those with very high narrowband excess ($>1.4$ mag or EW~$\gtrsim2000$~\AA\ in Fig.~\ref{fig6}), it seems likely that they are emitters other than PBEs, e.g., H$\alpha$ emitters at $z=5.2$ given the redshift evolution of EW$_\mathrm{H\alpha}$ ($\sim3700$~\AA\ at $z=5.2$ according to \citealt[fig.~11]{Rinaldi2023}).
However, we stress that 7 of 8 NBEs with such high narrowband excesses ($>1.4$ mag) are excluded by the color--color selection in the following.
Fig.~\ref{fig7} also shows that these emission line contributions are expected to occur solely in star-forming galaxies and AGNs, although quiescent galaxies with small PAH features at $z=0.2$ may also appear in the narrowband selection. 
Moreover, it indicates that the general color term between the F405N and F410M filters is negligible because the offset in their center wavelengths is small (0.037 $\mu$m, Fig.~\ref{fig1}).

\begin{figure*}
\centering
\includegraphics[width=0.9\textwidth]{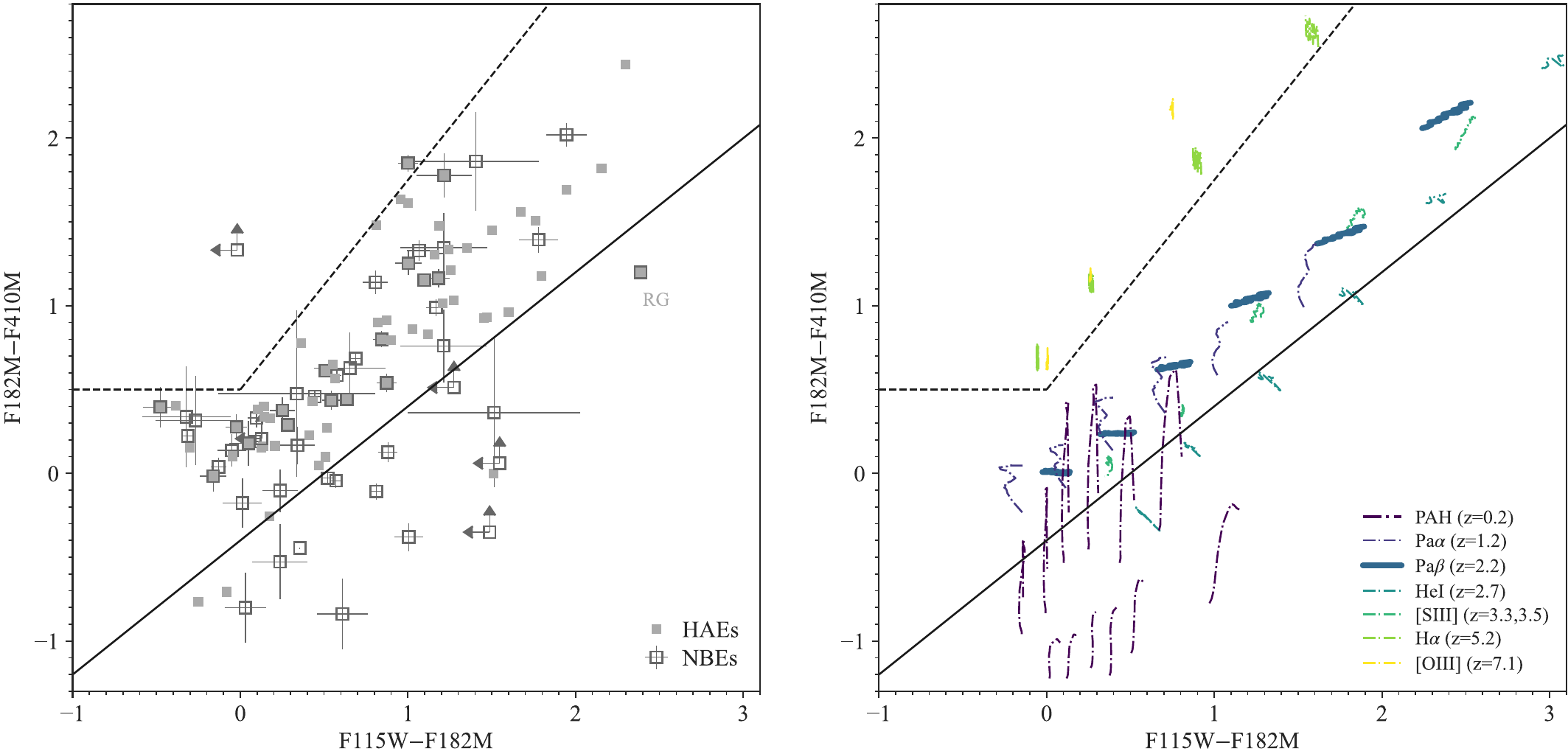}
\caption{
Color--color diagram (F115W$-$F182M versus F182M$-$F410M) for NBEs on the left panel and model templates on the right panel.
Open and filled squares in the left show NBEs and known HAEs within the FoV \citep[table~A]{Shimakawa2024}, respectively.
All colors are based on fixed aperture photometry of 0.3 arcsec diameter and error-bars are calculated from $1\sigma$ photometric uncertainties.
The black solid and dashed lines indicate the color thresholds adopted in this work to select PBE candidates at $z=2.2$.
On the right panel, colored curves depict the model colors based on SED templates from {\sc eazy} \citep{Brammer2008}: all 12 templates for PAH~$\lambda3.3\mu$m line at $z=0.23\pm0.10$ in purple, and 6 templates of star-forming galaxies for the other line emitters (dark blue: Pa$\alpha$ at $z=1.17\pm0.10$, blue: Pa$\beta$ at $z=2.17\pm0.10$, cyan: He{\sc i} at $z=2.74\pm0.10$, green: [S{\sc iii}] at $z=3.36\pm0.20$, yellow-green: H$\alpha$ at $z=5.18\pm0.10$, and yellow: [O{\sc iii}] at $z=7.10\pm0.10$). 
}
\label{fig8}
\end{figure*}

In light of these trends, we apply a color--color selection based on the obtained medium and broad-band images (F115W$-$F182M vs. F182M$-$F410M), excluding foreground and background contaminants on best-effort basis.
Fig.~\ref{fig8} shows NBEs (Section~\ref{s31}) on the color--color diagram on the left side, and the expected locations of line emitters at various redshifts possibly chosen by narrowband selection on the right side. 
We correct Galactic extinction by assuming the dust reddening value of $E(B-V)=0.0336$ \citep{Schlafly2011} and following the \citet{Cardelli1989} extinction curve.
On the right panel of Fig.~\ref{fig8}, we adopt model templates at $z_\mathrm{center}\pm0.1$ from {\sc eazy} \citep{Brammer2008}, but only those of star-forming galaxies except for the PAH$\lambda3.3\mu$m line at $z=0.2$ (see discussion above), where $z_\mathrm{center}$ indicates the emission-line redshift caught by the center wavelength of the F405N narrowband filter.
Fig.~\ref{fig8} also demonstrates that the color distributions of the model templates at $z=2.2$ are consistent with those of known protocluster members (HAEs) at the same redshift.

Based on the known HAEs and model templates, we set the following color thresholds (see also Fig.~\ref{fig8}),
\begin{eqnarray}
\mathrm{F182M}-\mathrm{F410M} &>& 0.8(\mathrm{F115W}-\mathrm{F182M}) -0.4, \label{eq7}
\end{eqnarray}
and
\begin{eqnarray}
\mathrm{F182M}-\mathrm{F410M} &<& 0.5, ~~ \mathrm{or} \label{eq8} \\
\mathrm{F182M}-\mathrm{F410M} &<& 1.25(\mathrm{F115W}-\mathrm{F182M}) +0.5. \label{eq9}
\end{eqnarray}
These color criteria enclose most of the NIRCam sources with HAE counterparts within 0.5 arcsec (63 out of 65 with duplicates) and 39 out of 57 NBEs. 
Additionally, we include two omitted NBEs in the final PBE candidates because they have HAE counterparts (Fig.~\ref{fig8}). 
One should note that two NBEs without F115W and F182M detection are not shown in the figure, which are also removed from the PBE candidates.
However, as inferred from the SED templates, completely removing potential foreground/background contaminants is challenging with only two color information because the color distribution of PBEs largely overlaps with those of some foreground/background emitters, particularly $z=0.2$ PAH, $z=1.2$ Pa$\alpha$, and $z=3.4$ [S{\sc iii}] emitters (Fig.~\ref{fig8}). 
Meanwhile, we cannot rule out the possibility that excluded sources, especially those without F182M detection at the lower right of the figure, could be PBEs.
Besides, the color--color diagram suggests detection of a HAE at $z=5.2$ or [O{\sc iii}] emitter at $z\sim7.1$, showing a significantly redder F182M$-$F410M color than the other NBEs for a given F115W$-$F182M color.
However, if we apply the recent measurement of H$\alpha$ luminosity at $z=6$ \citep[fig.~8]{Sun2023} to the survey field, it is estimated that $\sim6$ H$\alpha$ emitters may be detected. 
Furthermore, \citet[fig.~13]{Wold2024} have constrained [O{\sc iii}] luminosity function at $z=7$, suggesting that we may find 0--2 [O{\sc iii}] emitters.
Thus, as only one candidate in the background is lower than what we roughly expected, there could be some background emitters that have not yet been removed in the PBE candidates.

\subsection{Emitter catalog} \label{s33}

Based on our analyses, we select 57 NBEs using narrowband selection and exclude 16 NBEs via color--color selection from PBE candidates (Section~\ref{s32}). 
Thus, we have narrowed the original NBE samples down to 41 PBE candidates likely to be associated with the Spiderweb protocluster at $z=2.16$, which meet the following selection criteria:
\begin{eqnarray}
\mathrm{ID_{S18}} &\neq& \varnothing, ~~ \mathrm{or} \label{eq10} \\
\mathrm{Flag_{color}} &=& 1, \label{eq11}
\end{eqnarray}
where $\mathrm{ID_{S18}}$ is the identification number of the HAE counterparts in \citet{Shimakawa2018b,Shimakawa2024}, and $\mathrm{Flag_{color}}$ is the color selection flag in Section~\ref{s32}. 
The color selection flag denotes $\mathrm{Flag_{color}}=1$ if the F115W$-$F182M and F182M$-$F410M colors fit the selection criteria (Eq.~\ref{eq7}-\ref{eq9}).

\begin{table}
\centering
\caption{
List of the NBE samples (Section~\ref{s33}). 
Full table is attached at the end of this paper (Table~\ref{tab2}).
}
\label{tab1}
\begin{tabular}{ccccc}
\hline
ID & ID$_\mathrm{S18}$ & Flag$_\mathrm{color}$ & $F_\mathrm{NB}$ & EW$_{z=2.16}$\\
& & & ($10^{-18}$ erg~s$^{-1}$cm$^{-2}$) & (\AA)\\
\hline
    1 & --- & 1 & $16.0\pm2.7$ & $145\pm7$ \\
    2 & --- & 1 & $11.5\pm4.7$ & $160\pm17$ \\
    3 & --- & 1 & $3.0\pm2.1$  & $68\pm22$ \\
    \multicolumn{5}{c}{...}\\
\hline
\end{tabular}
\end{table}

The object identification numbers, color selection flags, line fluxes, and the rest-frame EWs at $z=2.16$ (EW$_{z=2.16}$) for all 57 NBEs are summarized in Table~\ref{tab1}.
We estimate the total NB fluxes and EW$_{z=2.16}$ for individual NBEs via the following prescription:
\begin{eqnarray}
F_\mathrm{NB} &=& \frac{\Delta_\mathrm{NB}(f_\mathrm{NB}-f_\mathrm{BB})}{1-\Delta_\mathrm{NB}/\Delta_\mathrm{BB}},\label{eq12}\\
\mathrm{EW}_{z=2.16} &=& \frac{\Delta_\mathrm{NB}(f_\mathrm{NB}-f_\mathrm{BB})}{f_\mathrm{BB}-f_\mathrm{NB}\cdot\Delta_\mathrm{NB}/\Delta_\mathrm{BB}}\frac{1}{3.16},\label{eq13}
\end{eqnarray}
where $\Delta$ is the filter bandwidth ($\Delta_\mathrm{NB}=460$~\AA\ and $\Delta_\mathrm{BB}=4360$~\AA). 
The total flux densities in the F405N ($f_\mathrm{NB}$) and F410M ($f_\mathrm{BB}$) bands are based on {\tt MAG\_AUTO} photometry in the {\sc SExtractor} \citep{Bertin1996}, and their photometric errors are calculated via Eq.~\ref{eq3}-\ref{eq4}.
We do not consider potential filter loss due to the small difference between the two filter throughput since most Pa$\beta$ line wavelengths of HAEs with spectroscopic redshifts are well covered by the N405N filter.


\section{Discussion} \label{s4}

\subsection{Spatial distribution} \label{s41}

\begin{figure*}
\centering
\includegraphics[width=0.93\textwidth]{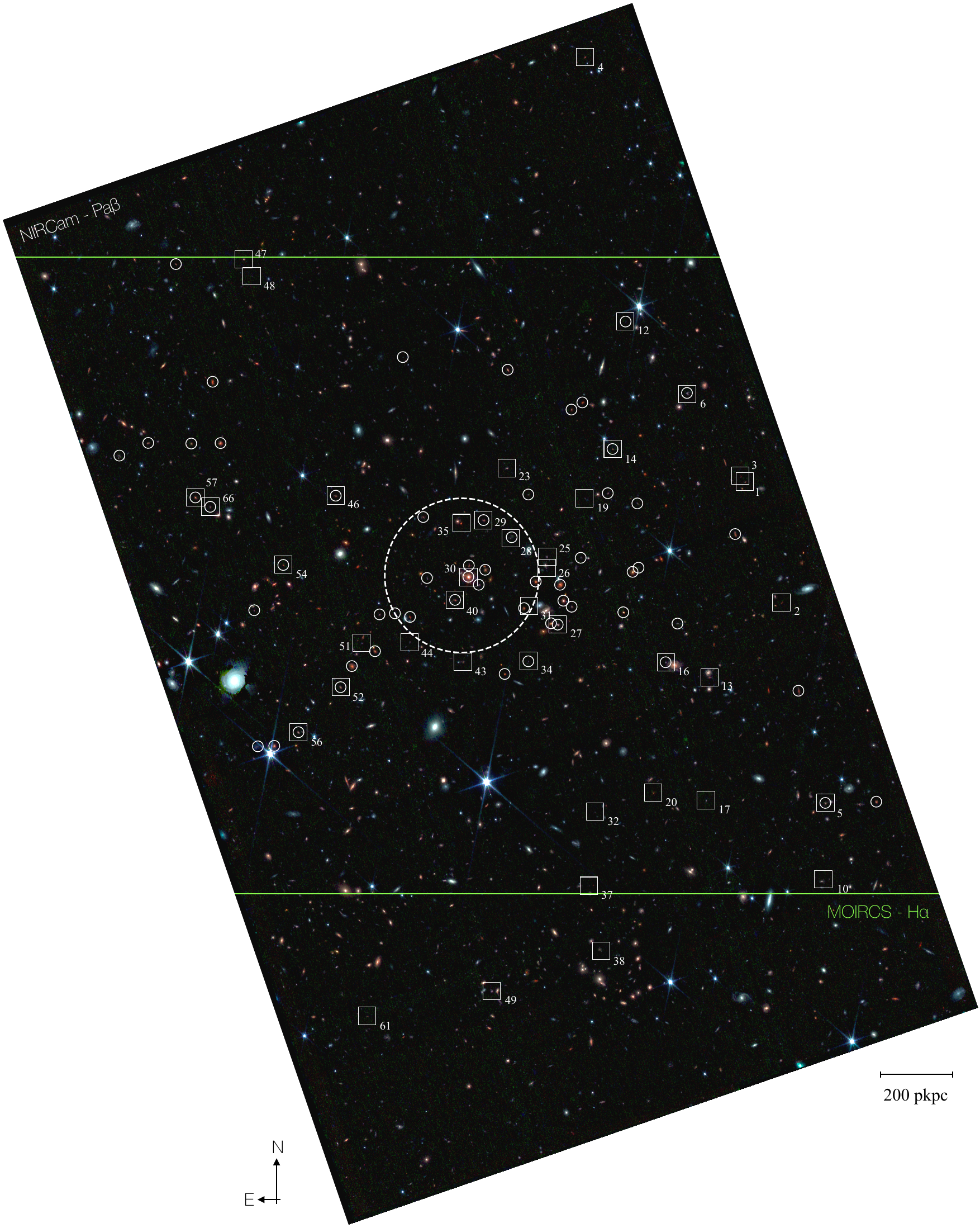}
\caption{
Spatial distribution of PBE candidates (squares) and known HAEs (circles; \citealt[table~A]{Shimakawa2024}) around the Spiderweb radio galaxy on the RGB (F410M/F182M/F115W) image made by {\sc stiff} \citep{Bertin2012}.
We here plot PBE candidates that satisfy the selection criteria (Eq.~\ref{eq10}-\ref{eq11}).
The green lines depict the survey area of the H$\alpha$ line imaging.
The dashed circle indicates the virial radius ($r_{500}=220$ proper kpc) based on the X-ray measurement by \citet{Tozzi2022b}. 
}
\label{fig9}
\end{figure*}

The spatial distribution of 41 PBE candidates and 58 known HAEs around the Spiderweb radio galaxy is shown in Fig.~\ref{fig9}; both samples are plotted on the RGB color image (F410M/F182M/F115W) covered by the JWST/NIRCam imaging.
Notably, many of the known HAE samples (41/58) are not selected as PBEs owing to the selection effects discussed in Section~\ref{s31}.
Therefore, we observe an apparent segregation between HAEs and PBE candidates in spatial distributions on the local scale. 
The small offset of redshift coverage by H$\alpha$ and Pa$\beta$ narrowband filters is another causal factor of the difference in spatial distribution. 
However, their distribution patterns are broadly consistent with each other; both populations spread towards the east and west from the center, while only a few galaxies are located in the northern and southern areas (Fig.~\ref{fig9}).
Some PBE candidates without HAE counterparts (\#10, 17, 20, 32, 37 and \#38, 49, 61 outside of the H$\alpha$ imaging) could form a substructure in the southern region or the bottom part of Fig.~\ref{fig9}, where no HAE have been detected in \citet{Koyama2013a,Shimakawa2018b}.
They could trace a backward structure at $z\sim2.18$ in the Spiderweb protocluster undetectable by the H$\alpha$ narrowband filter (at $z>2.175$, Fig.~\ref{fig1}). 
However, we cannot rule out the possibility of projection effects due to non-Pa$\beta$ emitters in the foregrounds and/or backgrounds.

We now focus more on some individual PBE candidates.
Fig.~\ref{fig9} shows two new member candidates (\#31 and \#35) within $r_{500}$ radius of 220 proper kpc \citep{Tozzi2022b}. 
\#31 is a faint source that is marginally detected in the NIRCam images but invisible in the currently available ground-based data.
We consider that \#35 might have been missed in previous H$\alpha$ imaging \citep{Kurk2004a,Koyama2013a,Shimakawa2018b} because it was disturbed by a nearby bright object at a distance of 1.3 arcsec in the seeing-limited narrowband image.
Notably, another new candidate deblended by the NIRCam high-resolution images is observed (\#26 located 1 arcsec from a brighter source).
Conversely, as summarized in Table~\ref{tab1}, \#5, 6, 12, 14, 16, 27, 28, 29, 30, 34, 40, 46, 52, 54, 56, 57, 66 have HAE counterparts in \citet{Shimakawa2018b}; hence, they are so-called dual H$\alpha$ + Pa$\beta$ emitters. 
Their emission line ratios and environmental dependence are discussed by P\'{e}rez-Mart\'{\i}nez et al. (2024).
Cross-matching with other populations, such as quiescent galaxies and submm sources, and detailed analyses of their physical properties are left to future studies.

\subsection{Line luminosity function} \label{s42}

The line luminosity function gives fundamental information regarding galaxy abundance as a function of luminosity \citep{Sobral2010,Sobral2011,Sobral2013,Sobral2015,Hayes2010,Ly2011,Matthee2017,Shimakawa2018a,Hayashi2018,Hayashi2020,Zheng2021,Nagaraj2023}. 
This section estimates Pa$\beta$ line luminosity function based on the selected PBE candidates.
The number excess of massive star-forming galaxies (or AGNs) in the Spiderweb protocluster compared to the general field has been reported in previous studies \citep{Hatch2011,Koyama2013a,Koyama2013b,Shimakawa2018b}, which would enhance the line luminosity function at the bright end. 
In contrast, narrowband selection only selects Pa$\beta$ emitters with high EWs ($>20$ \AA) and miss massive galaxies, as explained in Section~\ref{s31}, which would suppress the number density at the bright end.

We calculate the number densities per unit Pa$\beta$ luminosity following the $V_\mathrm{max}$ procedure,
\begin{equation}
\phi_\mathrm{obs}(\log{L}) = \sum_i\frac{1}{V_\mathrm{max}C\Delta(\log{L})}, \label{eq14}
\end{equation}
where $V_\mathrm{max}$ is the survey volume ($=1812$ co-Mpc$^3$) of the NIRCam Pa$\beta$ narrowband imaging (Fig.~\ref{fig2}), and $C$ is the selection completeness obtained through the Monte Carlo simulation (Section~\ref{s31}). 
The survey volume is determined from the survey area (20.90 arcmin$^2$ or 53.91 co-Mpc$^2$) and the filter width of the F405N narrowband filter (0.046 $\mu$m or 33.61 co-Mpc).
We do not apply the completeness correction to the bright-end sources at F405N $<22$ mag (applying to only the radio galaxy) since the simulated completeness value for typical NBEs is nearly zero (Fig.~\ref{fig6}).
Notably, dust attenuation is not considered in this work because it is discussed in the companion study (P\'{e}rez-Mart\'{\i}nez et al. 2024), and PBE candidates are not fully covered by the previous H$\alpha$ narrowband imaging (Fig.~\ref{fig2}).
Given the limited sample size, we adopt a relatively large bin size of $\Delta(\log{L})=0.2$.

\begin{figure}
\centering
\includegraphics[width=8.5cm]{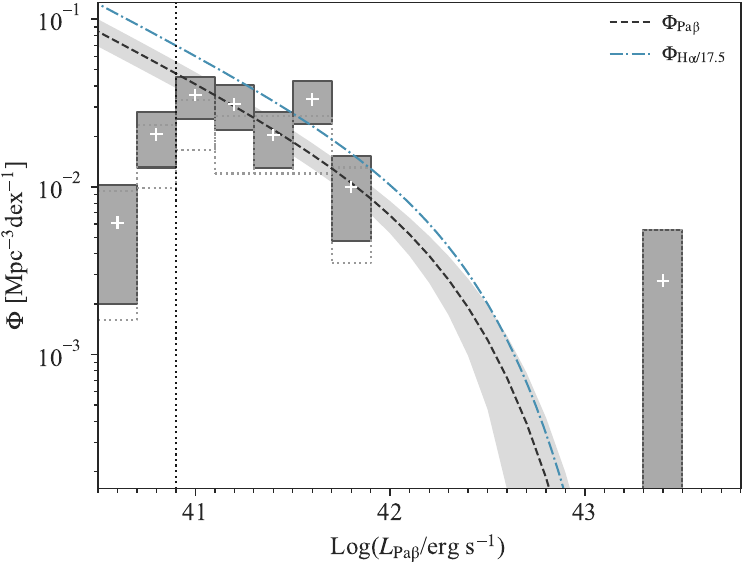}
\caption{
Pa$\beta$ luminosity function for the selected PBE candidates in the Spiderweb protocluster at $z=2.16$.
The gray filled and open dotted squares show the number densities in each Pa$\beta$ luminosity bin with and without the completeness correction (Section~\ref{s31}), respectively.
The best-fit Pa$\beta$ luminosity function is depicted by the black dashed curve with a $1\sigma$ error range in gray, where the density bins below the threshold marked by the dotted vertical line is not taken into account.
The blue dash-dotted curve is the dust-corrected H$\alpha$ luminosity function for the HAE sample reported by \citet{Shimakawa2018b}, including a scaling factor of H$\alpha$/Pa$\beta=17.5$.
}
\label{fig10}
\end{figure}

The resultant Pa$\beta$ luminosity function is shown in Fig.~\ref{fig10}.
The luminosity function can be formed by a Schechter function \citep[$\phi(L)$]{Schechter1976}, given by the following equation:
\begin{equation}
\phi(L)d(\log{L}) = \phi^\ast\left(\frac{L}{L^\ast}\right)^{\alpha+1}\exp\left(-\frac{L}{L^\ast}\right)\ln{10}~d(\log{L}). \label{eq15}
\end{equation}
This is a logarithmic expression transformed from the original formula. 
$\phi^\ast$, $L^\ast$, and $\alpha$ are Schechter parameters, where $\phi^\ast$ is a normalization density at a characteristic luminosity $L^\ast$, and $\alpha$ is a scaling exponent of the power law driving the luminosity function at the faint end ($L\ll L^\ast$).
We perform a chi-square fitting to the $\phi_\mathrm{obs}(\log{L})$ values (Eq.~\ref{eq14}) above the luminosity threshold of $L_\mathrm{H\alpha,corr}=10^{40.9}$ erg~s$^{-1}$ using a fitting algorithm, {\sc lmfit} \citep[version~1.2.1]{Newville2023}.
We do not consider luminosity bins below $\log(L\mathrm{_{Pa\beta}/erg~s^{-1}})=40.9$ (F405N $\sim25$ mag) during curve fitting because we can not well recover the number densities due to the severe lack of the selection completeness below $\sim50$\% (Fig.~\ref{fig6}).
Consequently, we obtain the best-fit parameters $\log{\phi^\ast}=-2.53\pm0.15$ and $\log{L^\ast}=42.33\pm0.17$, as indicated in Fig.~\ref{fig10}.
Here, given the limited sample size, the scaling exponent is fixed to $\alpha=-1.6$ according to the studies by \citet{Sobral2013,Hayashi2018}, which reported that the faint-end slope is reasonably fitted by $\alpha=-1.6$ regardless of redshift at least up to $z\sim2.2$.
We confirm that the fitting result is not significantly improved even if we choose different $\alpha$ values. 
The effects of stellar absorption are ignored, assuming they are sufficiently small throughout the study (e.g., EW$\lesssim$ 1--2~\AA\ in absorption according to \citealt{Calabro2019,Seille2024}), owing to the lack of observational constraints. 
Moreover, the non-tophat transmittance profile of the F405N narrowband filter could affect the shape of Pa$\beta$ luminosity function at the bright end \citep{Sobral2013,Shimakawa2018a}.
However, as we have not been able to constrain the bright-end densities of PBEs, we do not consider such an effect.

The best-fit Pa$\beta$ luminosity function of the PBE candidates is 1.5 times lower than the dust-corrected H$\alpha$ luminosity function by \citet{Shimakawa2018b} assuming the H$\alpha$/Pa$\beta$ line ratio of 17.5.
The differential is presumably due to the selection bias in the narrowband selection, given the significant fraction of unmatched samples between PBE candidates and HAEs. 
Particularly, no Pa$\beta$-luminous emitters with $L\mathrm{_{Pa\beta}}>10^{42}$ erg~s$^{-1}$ are detected, with the exception of the Spiderweb radio galaxy (\#30). 
The different survey areas (Fig.~\ref{fig2}) also contribute to the difference since the H$\alpha$ imaging is more optimized to dense substructures in the protocluster.
In addition, the dust extinction would have a minor effect, which increases Pa$\beta$ luminosity by 0.13 dex when assuming Pa$\beta$ extinction of 0.33 mag (or $A_\mathrm{H\alpha}=1$; P\'{e}rez-Mart\'{\i}nez et al. 2024).
Besides, \citet{Shimakawa2018b} have reported an overdensity of $\delta=12.29$ in the H$\alpha$ luminosity function when compared to that in the general field \citep{Sobral2014}.
From these comparisons, a number excess of the obtained Pa$\beta$ luminosity function is tentatively estimated to be $\delta\sim8$. 
However, one should note that the current Pa$\beta$ luminosity function needs to be considered as an upper limit because it would contain other line emitters at $z\ne2.2$ (Section~\ref{s32}).

\subsection{Possible emitters seen only in the narrowband} \label{s43}

Finally, we discuss the possibility of missing faint emitters visible only in the F405N image, hereafter referred to as NB-only emitters.
Since we use the median-stacked image of all four band filters for source detection (Section~\ref{s22}), we could overlook emission-line dominated emitters that are too faint to be detected in the F115W/F182M/F410M bands and, hence, in the median combined image.
Such peculiar sources have been observed commonly in the narrowband imaging survey (e.g., \citealt{Hayashi2016,Shibuya2018}).
Although their characteristics largely depend on the limiting magnitudes of the observing bands and their combination, they generally tend to be low-mass star-forming galaxies with high specific star formation rates (e.g. \citealt{Hayashi2016}) like extreme emission-line galaxies \citep{Cardamone2009,Amorin2010,Izotov2012,Kojima2020}.

\begin{figure}
\centering
\includegraphics[width=7.5cm]{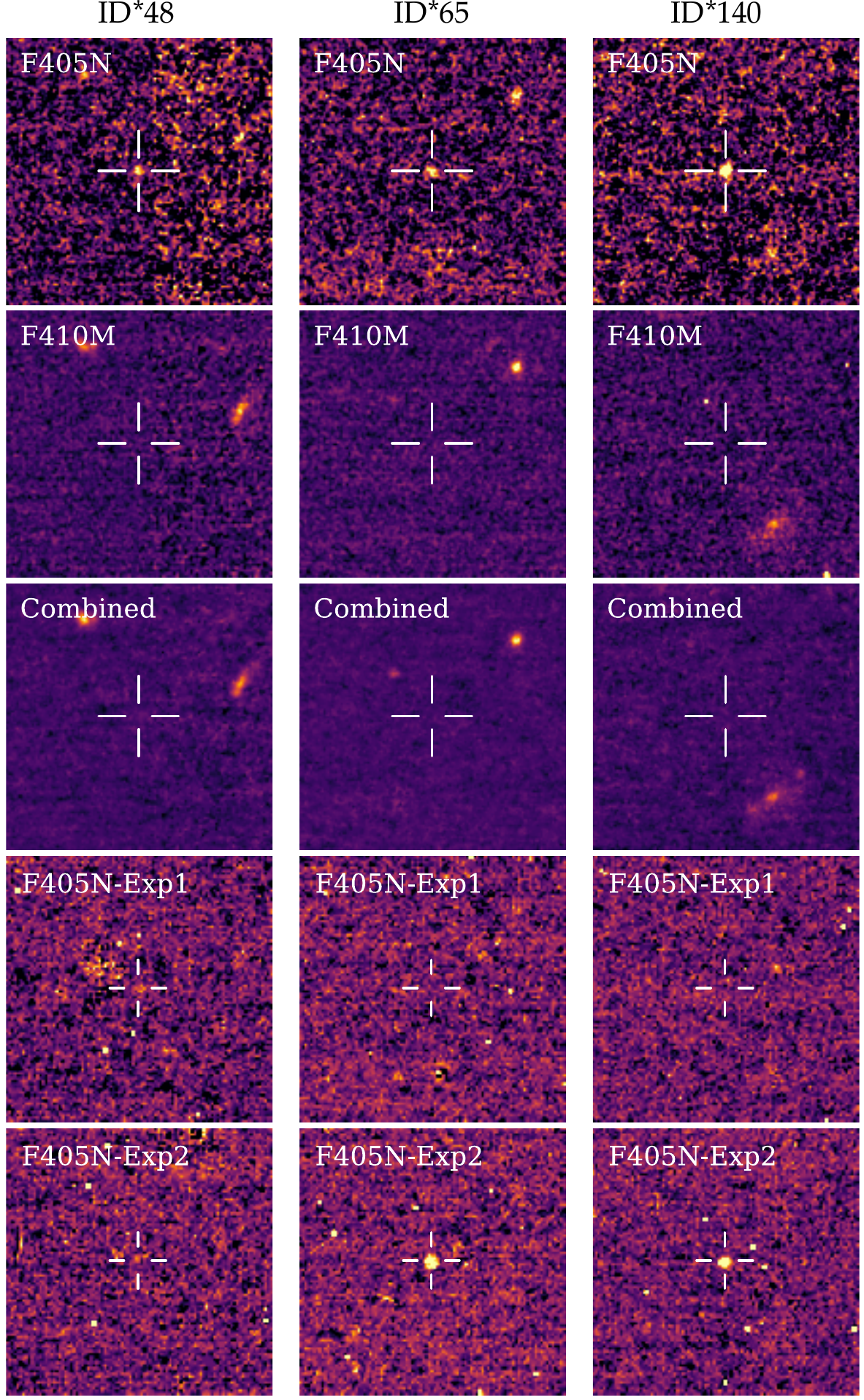}
\caption{
Example cutouts of 
NB-only emitters that are not used in this study.
The first, second, and third rows show them on the F405N, F410M, and 4-band combined images, respectively.
They are normalized with a linear stretch in the same flux range. 
The forth and fifth rows show individual exposures in the F405N band for each source, which are normalized in a different range to the first-to-third rows for the sake of visibility.
For the two sources on the right and middle panels, cosmic rays are detected in one of the exposures.
}
\label{fig11}
\end{figure}

Therefore, we rerun the narrowband selection as in Section~\ref{s31}, but we adopt the F405N image as the detection image instead of the combined image.
Here, we modify a parameter of {\tt DETECT\_MINAREA} $=15$ and add a Gaussian filter of {\tt gauss\_3.0\_5x5} to capture very faint objects, although these parameter changes significantly increase false detection. 
Then, we select additional NBE candidates satisfying the selection criteria (Eq.~\ref{eq5}-\ref{eq6}) from the undetected samples in the F410M band above two sigma levels, amounting to 138 sources. 
In the end, however, only 26 candidates remain after a careful visual inspection.
The excluded samples comprise obvious artifacts due to cosmic rays with one or a few bright pixels.
Fig.~\ref{fig11} shows some examples of the screened sources, including how they are present only in the F405N image and not detected in the F410M and combined images.

\begin{figure}
\centering
\includegraphics[width=7.5cm]{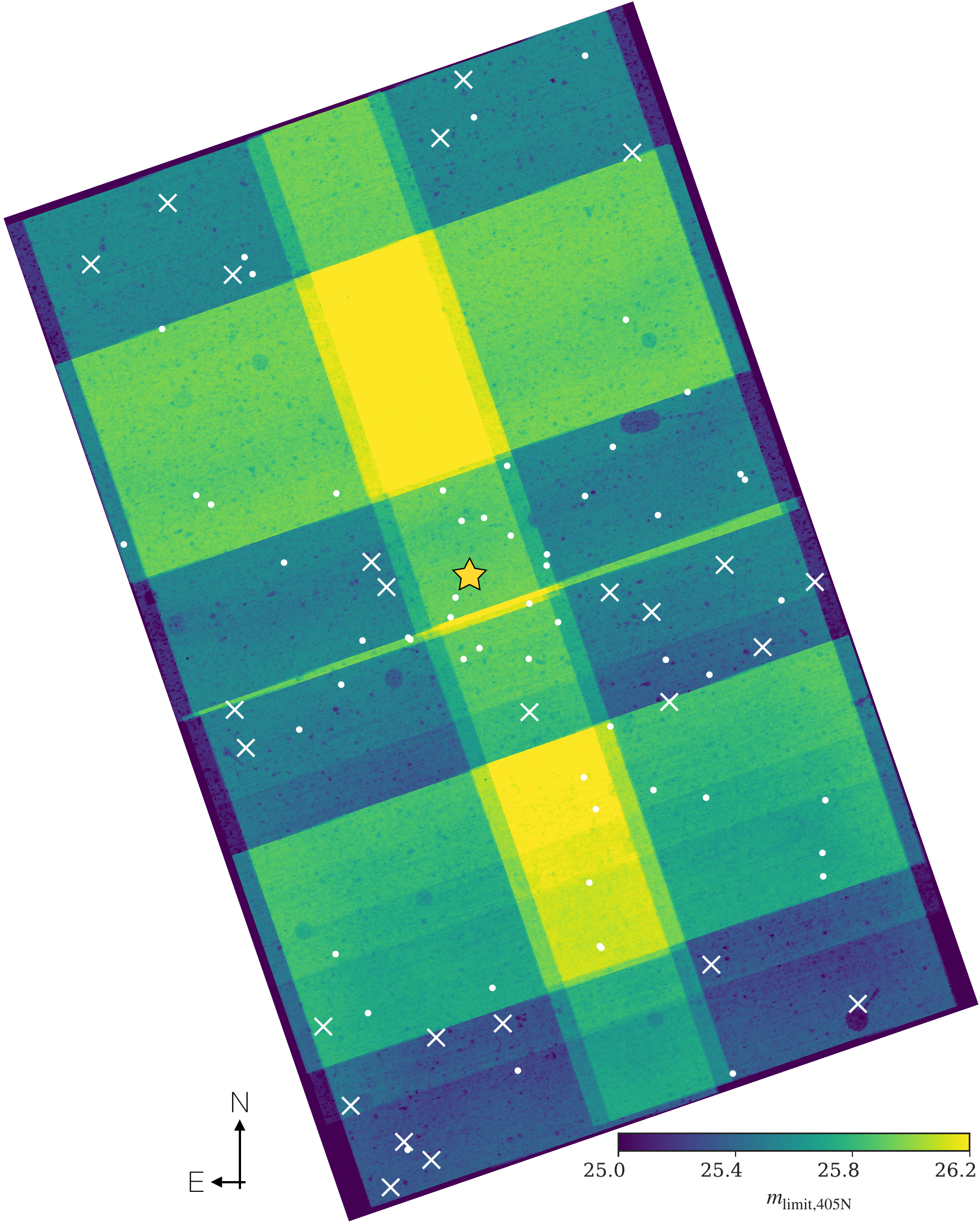}
\caption{
Spatial distribution of NBE candidates with only the narrowband detection (white crosses). 
The color map in the background shows the depth in the F405N band as shown in Fig.~\ref{fig2}.
They tend to be located in the shallowest field with the number of exposures $N_\mathrm{exp}=2$.
The small white circles are NBEs selected in the main analysis (Section~\ref{s31}).
The yellow star is the Spiderweb radio galaxy.
}
\label{fig12}
\end{figure}

We examine their spatial distribution on the weight map in the F405N band for further validation, as shown in Fig.~\ref{fig12}.
Notably, most candidates of NB-only emitters are distributed in areas of a small number of exposures (corresponding to lower weight values), making the credibility of these detections highly questionable.
As these shallow areas comprise only two integrations, some artifacts most likely remain owing to the inability to measure the median values when combining the images. 
In fact, we confirm that 88\% (23/26) of them are most likely surviving cosmic rays by checking individual exposures as illustrated in Fig.~\ref{fig11}. 
Although remaining three sources could be real objects, they can be explained by a result of random variation in three sigma threshold in our narrowband selection (Eq.~\ref{eq5}) as the total number of source detection used here is $\sim2700$.
Considering the results of these validation checks, we decide not to include these emitter candidates in the main targets, and mention them in this last section instead of the main text.

\section{Conclusions} \label{s5}

We present our initial results on Pa$\beta$ line imaging with JWST/NIRCam towards one of the best-studied protoclusters known as the Spiderweb protocluster at $z=2.16$ \citep{Carilli1997,Pentericci1997,Pentericci2000,Kurk2000}. 
To maximize the science outputs, we focus on the narrowband F405N and medium-band F410M imaging, and in parallel, the F115W and F182M imaging on the blue channel, allowing us to obtain Pa$\beta$ line and the rest-frame $UVJ$ colors of protocluster members across the survey field.
This paper summarizes the narrowband technique and color--color selection to find candidates of Pa$\beta$ emitters (PBEs) as one of the first reports in the series.
Then, we investigate the Pa$\beta$ luminosity function by correcting the selection completeness.

The research highlights of this study are described as follows: We first conduct narrowband selection on the color--magnitude diagram (F410M$-$F405N versus F405N) to select narrowband emitters (NBEs).
Consequently, we obtain 57 NBEs showing flux excesses in the F405N band above three sigma levels ($\Sigma>3$). 
Subsequent color--color selection (F182M$-$F410M versus F115W$-$F182M) helps us remove potential foreground or background contaminants, such as PAH$\lambda3.3\mu$m at $z=0.2$ and H$\alpha$ emitters at $z=5.2$, further narrowing the PBE candidates down to 41 sources.
As 17 of the PBE candidates are also confirmed as H$\alpha$ emitters based on the previous H$\alpha$ imaging \citep{Shimakawa2018b}, the remaining 24 objects are considered to be unconfirmed candidates associated with the Spiderweb protocluster.
These PBE candidates are still at risk of foreground or background emitters other than PBEs; therefore, further follow-up studies are needed to establish that they are protocluster members.
Additionally, one H$\alpha$ or [O{\sc iii}] emitter candidate at $z>5$ is obtained via the selection process.

The original goal of this study was to make a deep line survey aimed at PBEs with a unique narrowband filter on JWST/NIRCam, taking advantage of NIR observation less sensitive to the dust extinction.
In fact, we have succeeded in finding new candidates of protocluster members. 
However, the general narrowband technique requires a relatively high EW threshold, which makes it difficult to select bright (and dust-obscured) PBEs as demonstrated in this study.
In terms of achieving more unbiased search of NIR lines with low EWs ($\lesssim20$ \AA), the wide field slitless spectroscopy would be more suitable than the narrow-band imaging.

The final PBE candidates selected in this study show a similar spatial distribution of HAEs, as established in the previous studies \citep{Koyama2013a,Shimakawa2018b}.
We then investigate the Pa$\beta$ luminosity function for the PBE candidates by measuring their Pa$\beta$ fluxes based on the newly obtained JWST/NIRCam imaging data. 
The characteristic Pa$\beta$ luminosity and normalization density are estimated to be $\log{L^\ast}=42.33\pm0.17$ and $\log{\phi^\ast}=-2.53\pm0.15$, respectively.
Follow-up confirmations and characterizations of the PBE candidates will provide a better understanding of the total star formation rate in the Spiderweb protocluster, environmental dependence of galaxy formation, and a transition process from a protocluster to a bonafide cluster of galaxies.



\section{Acknowledgments}

We thank the anonymous referee for useful comments. 
This work is based on observations made with the NASA/ESA/CSA James Webb Space Telescope. 
The data were obtained from the Barbara A. Mikulski Archive for Space Telescopes at the Space Telescope Science Institute\footnote{\dataset[10.17909/vx25-q902]{\doi{10.17909/vx25-q902}}}, which is operated by the Association of Universities for Research in Astronomy, Inc., under NASA contract NAS 5-03127 for JWST. 
The observation is associated with program \#1572 in cycle~1.

This work is supported by a Waseda University Grant for Special Research Projects (2023C-590) and MEXT/JSPS KAKENHI Grant Number (23H01219).
We would like to thank Editage (www.editage.com) for English language editing. 
JMPM acknowledges funding from the European Union’s Horizon Europe research and innovation programme under the Marie Sk{\l}odowska-Curie grant agreement No 101106626.
HD, JMPM and YZ acknowledge financial support from the Agencia Estatal de Investigaci\'on del Ministerio de Ciencia e Innovaci\'on (AEI-MCINN) under grant (La evolución de los c\'umulos de galaxias desde el amanecer hasta el mediod\'ia c\'osmico) with reference (PID2019-105776GB-I00/DOI:10.13039/501100011033) and Agencia Estatal de Investigaci\'on del Ministerio de Ciencia, Innovaci\'on y Universidades (MCIU/AEI) under grant (Construcci\'on de c\'umulos de galaxias en formaci\'on a trav\'es de la formaci\'on estelar oscurecida por el polvo) and the European Regional Development Fund (ERDF) with reference (PID2022-143243NB-I00/10.13039/501100011033).
TK acknowledges financial support from JSPS KAKENHI Grant Numbers 24H00002 (Specially Promoted Research by T. Kodama et al.) and 22K21349 (International Leading Research by S. Miyazaki et al.).
PGP-G acknowledges support from grant PID2022-139567NB-I00 funded by Spanish Ministerio de Ciencia y Universidades MCIU/AEI/10.13039/501100011033, FEDER {\it Una manera de hacer Europa}.
YZ acknowledges the support from the China Scholarship Council (202206340048), and the National Science Foundation of Jiangsu Province (BK20231106).
C.D.E. acknowledges funding from the MCIN/AEI (Spain) and the NextGenerationEU/PRTR (European Union) through the Juan de la Cierva-Formaci\'on program (FJC2021-047307-I)



%

\vspace{5mm}
\facility{JWST (NIRCam)}


\software{Numpy \citep{Harris2020},
          Pandas \citep{Reback2022},
          Matplotlib \citep{Hunter2007},
          Astropy \citep{AstropyCollaboration2013,AstropyCollaboration2018},
          Topcat \citep{Taylor2005},
          Source Extractor \citep{Bertin1996},
          Photutils \citep{Bradley2023},
          Lmfit \citep{Newville2023},
          Stiff \citep{Bertin2012}
          }


\bibliography{rs23b}{}
\bibliographystyle{aasjournal}



\begin{table}
\centering
\caption{
Full table of the NBE samples (Section~\ref{s33}). 
}
\label{tab2}
\begin{tabular}{ccccc}
\hline
ID & ID$_\mathrm{S18}$ & Flag$_\mathrm{color}$ & $F_\mathrm{NB}$ & EW$_{z=2.16}$\\
& & & ($10^{-18}$ erg~s$^{-1}$cm$^{-2}$) & (\AA)\\
\hline
1  & --- & 1 & $16.0\pm2.7$ & $145\pm7$ \\
2  & --- & 1 & $11.5\pm4.7$ & $160\pm17$ \\
3  & --- & 1 &  $3.0\pm2.1$ & $68\pm22$ \\
4  & --- & 1 &  $4.1\pm3.7$ & $22\pm15$ \\
5  &  11 & 1 &  $5.4\pm6.5$ & $31\pm26$ \\
6  &  84 & 1 &  $5.5\pm1.2$ & $475\pm10$ \\
8  & --- & 0 &  $2.2\pm3.1$ & $22\pm24$ \\
10 & --- & 1 &  $5.4\pm2.5$ & $961\pm18$ \\
11 & --- & 0 & $12.1\pm8.9$ & $54\pm22$ \\
12 &  96 & 1 &  $2.6\pm2.1$ & $164\pm34$ \\
13 & --- & 1 & $13.4\pm6.2$ & $61\pm14$ \\
14 &  76 & 1 &  $3.4\pm3.7$ & $39\pm27$ \\
16 &  23 & 1 &  $7.5\pm3.6$ & $142\pm19$ \\
17 & --- & 1 & $16.9\pm3.0$ & $260\pm7$ \\
18 & --- & 0 &  $3.3\pm1.0$ & $>1000$ \\
19 & --- & 1 &  $2.9\pm2.5$ & $93\pm31$ \\
20 & --- & 1 &  $2.0\pm2.7$ & $16\pm17$ \\
22 & --- & 0 &  $6.5\pm2.0$ & $>1000$ \\
23 & --- & 1 &  $4.3\pm5.4$ & $20\pm19$ \\
24 & --- & 0 &  $6.0\pm2.6$ & $155\pm18$ \\
25 & --- & 1 &  $7.7\pm3.9$ & $67\pm17$ \\
26 & --- & 1 &  $1.6\pm1.0$ & $72\pm22$ \\
27 &  35 & 1 & $10.4\pm4.6$ & $27\pm8$ \\
28 &  59 & 1 &  $0.9\pm0.3$ & $49\pm10$ \\
29 &  65 & 1 & $19.2\pm5.0$ & $111\pm10$ \\
30 &  73 & 0 &  $843\pm14$  & $52\pm0$ \\
31 & --- & 1 &  $2.4\pm1.5$ & $339\pm26$ \\
32 & --- & 1 &  $2.8\pm1.4$ & $603\pm19$ \\
33 & --- & 0 &  $1.4\pm1.1$ & $341\pm34$ \\
34 &  24 & 1 &  $1.5\pm2.1$ & $69\pm46$ \\
35 & --- & 1 &  $2.0\pm2.0$ & $8\pm7$ \\
36 & --- & 0 &  $8.3\pm6.1$ & $83\pm26$ \\
37 & --- & 1 &  $1.7\pm1.5$ & $239\pm38$ \\
38 & --- & 1 &  $6.0\pm1.8$ & $512\pm12$ \\
39 & --- & 0 &  $3.5\pm1.3$ & $459\pm16$ \\
40 &  42 & 0 & $11.7\pm4.8$ & $37\pm10$ \\
41 & --- & 0 & $56.3\pm5.5$ & $107\pm4$ \\
42 & --- & 0 &  $1.8\pm1.0$ & $>1000$ \\
43 & --- & 1 &  $1.1\pm3.4$ & $10\pm26$ \\
44 & --- & 1 &  $6.5\pm2.5$ & $190\pm16$ \\
45 & --- & 0 &  $1.7\pm0.8$ & $>1000$ \\
46 &  69 & 1 &  $6.3\pm3.2$ & $41\pm13$ \\
47 & --- & 1 &  $4.3\pm4.6$ & $13\pm12$ \\
48 & --- & 1 &  $2.9\pm2.6$ & $333\pm39$ \\
49 & --- & 1 & $11.2\pm5.3$ & $82\pm16$ \\
50 & --- & 0 &  $2.0\pm0.5$ & $721\pm10$ \\
\hline
\end{tabular}
\end{table}

\begin{table}
\centering
\caption{
Continued. 
}
\begin{tabular}{ccccc}
\hline
ID & ID$_\mathrm{S18}$ & Flag$_\mathrm{color}$ & $F_\mathrm{NB}$ & EW$_{z=2.16}$\\
& & & ($10^{-18}$ erg~s$^{-1}$cm$^{-2}$) & (\AA)\\
\hline
51 & --- & 1 &  $8.9\pm3.7$ & $205\pm18$ \\
52 &  21 & 1 &  $2.3\pm2.3$ & $13\pm11$ \\
54 &  53 & 1 &  $2.1\pm3.2$ & $18\pm22$ \\
56 &  16 & 1 &  $8.1\pm4.8$ & $40\pm15$ \\
57 &  71 & 1 & $11.4\pm4.7$ & $31\pm9$ \\
59 & --- & 0 &  $3.6\pm0.9$ & $>1000$ \\
60 & --- & 0 &  $5.5\pm3.9$ & $103\pm27$ \\
61 & --- & 1 &  $3.3\pm2.5$ & $93\pm28$ \\
63 & --- & 0 &  $4.3\pm2.7$ & $510\pm26$ \\
65 & --- & 0 &  $2.2\pm2.0$ & $980\pm36$ \\
66 &  66 & 1 &  $3.6\pm3.5$ & $39\pm24$ \\
\hline
\end{tabular}
\end{table}

\end{document}